\documentclass{webofc}
\usepackage[varg]{txfonts}   
\newcommand{\be}{\begin{equation}}
\newcommand{\ee}{\end{equation}}
\newcommand{\bea}{\begin{eqnarray}}
\newcommand{\eea}{\end{eqnarray}}

\newcommand{\bml}{\begin{subequations}}
\newcommand{\eml}{\end{subequations}}

\newcommand{\bbm}{\begin{bmatrix}}
\newcommand{\ebm}{\end{bmatrix}}
\newcommand{\bvm}{\begin{vmatrix}}
\newcommand{\evm}{\end{vmatrix}}

\newcommand{\mc}{\mathcal}

\newcommand{\ve}{\varepsilon}
\newcommand{\n}{\nabla}

\newcommand{\cfe}{\tilde \ve}
\newcommand{\cfp}{\tilde \pi}
\newcommand{\cfn}{\tilde \nu}
\newcommand{\cfq}{\tilde \theta}
\newcommand{\cfj}{\tilde \gamma}

\begin{document}
\title{A new causal and stable theory of viscous chiral hydrodynamics}
%
%

\author{\firstname{Nick} \lastname{Abboud}\inst{1}\fnsep\thanks{\email{nka2@illinois.edu}} \and
        \firstname{Enrico} \lastname{Speranza}\inst{2}\fnsep\thanks{\email{enrico.speranza@cern.ch}} \and
        \firstname{Jorge} \lastname{Noronha}\inst{1}\fnsep\thanks{\email{jn0508@illinois.edu}}
}

\institute{Illinois Center for Advanced Studies of the Universe \& Department of Physics, \\
University of Illinois Urbana-Champaign, Urbana, IL 61801, USA
\and
Theoretical Physics Department, CERN, 1211 Geneva 23, Switzerland
}

\abstract{%
    We construct the general theory of first-order relativistic hydrodynamics for a fluid exhibiting a chiral anomaly, including all possible viscous terms allowed by symmetry. Using standard techniques, we compute the necessary and sufficient conditions for this theory to be relativistically causal in the nonlinear regime and for thermal equilibria to be linearly stable. This is the first theory of first-order chiral hydrodynamics suitable for numerical simulations.
}
\maketitle
\section{Introduction}
\label{intro}

Hydrodynamics is an effective description of many-body systems in the long-wavelength, long-time limit. An important topic of current research in the field is the extension of hydrodynamic theories to include quantum effects from the underlying theory describing interactions among the microscopic constituents. A prime example is chiral hydrodynamics, which incorporates the effect of quantum anomalies in the macroscopic fluid description \cite{Erdmenger:2008rm, Son:2009tf, Neiman:2010zi, Sadofyev:2010pr, Chen:2015gta}.
Theoretical, numerical, and experimental work is underway to estimate the strength of signatures of anomalous effects in the quark-gluon plasma (QGP) formed in heavy-ion collisions \cite{Kharzeev:2015znc}. Experimental evidence of anomalous phenomena in the QGP would shed light on CP-violation and the structure of the QCD vacuum \cite{Kharzeev:2007jp}.

Due to the importance of viscous effects in the QGP, it is crucial to formulate a theory of \emph{viscous chiral hydrodynamics} to study and simulate effects associated with quantum anomalies. However, strikingly, some of the previously developed theories are unsuitable for numerical simulation even in the ideal case \cite{Speranza:2021bxf}. Indeed, they are incompatible with two fundamental principles from relativity and thermodynamics---they violate relativistic causality, i.e., they support faster-than-light signal propagation, and they unphysically predict catastrophic instabilities of thermal equilibrium states. Some such theories do not even have a well-posed initial value problem; namely, they do not admit unique solutions for arbitrary initial data \cite{Speranza:2021bxf}.

A systematic effective field theory approach to relativistic hydrodynamics known as the BDNK method has recently been developed \cite{Bemfica:2017wps,Kovtun:2019hdm,Bemfica:2019knx,Hoult:2020eho,Bemfica:2020zjp}. In the BDNK method, all possible first-order terms allowed by symmetry are included in the constitutive relations of the hydrodynamic currents. Conditions for causality and stability are then established, constraining the physically permissible values of the various transport parameters. We apply the BDNK method to construct the first causal and stable formulation of viscous chiral hydrodynamics incorporating the chiral anomaly \cite{Abboud:2023hos}. This enables, for the first time, consistent numerical simulations of chiral transport effects throughout the hydrodynamic evolution of the QGP.

\section{First-order chiral hydrodynamics}

We assume that the chirally anomalous fluid is characterized by an axial current $J^\mu_5$ associated with an anomalous $U(1)_A$ symmetry, as well as the energy-momentum tensor $T^{\mu\nu}$. For simplicity, we neglect any conserved vector $U(1)_V$ currents, e.g., baryon density, and we couple the fluid to an artificial background $U(1)_A$ gauge field\footnote{This is the so-called $U(1)_A$ model \cite{Son:2009tf}. The more phenomenologically relevant $U(1)_A \times U(1)_V$ model, with the electromagnetic $U(1)_V$ symmetry gauged instead, can be found in \cite{Abboud:2023hos}.}. The equations of motion are then
\begin{align} \label{conservation}
    \partial_\mu T^{\mu\nu} = F^{\nu\lambda}J_{5\lambda}\qquad \text{and} \qquad \partial_\mu J^\mu_5 = C E^\mu B_\mu,
\end{align}
where $C$ is the chiral anomaly coefficient and $E^\mu = F^{\mu\nu} u_\nu$ and $B^\mu = \frac{1}{2}\epsilon^{\mu\alpha\beta\gamma}u_\alpha F_{\beta\gamma}$ are external ``electric and magnetic fields'' associated with the $U(1)_A$ gauge field with field strength $F^{\mu\nu}$. We take the dynamical fields describing the fluid to be energy density $\varepsilon$, axial charge density $n_5$, and flow velocity $u^\mu$, and express $J^\mu_5$ and $T^{\mu\nu}$ with the most general symmetry-allowed terms up to first order in derivatives \cite{Abboud:2023hos}. The constitutive relations are thus given by
\begin{subequations}\label{decomp-kovtun}
\begin{align} 
    T^{\mu\nu} &= \mc Eu^\mu u^\nu + \mc P \Delta^{\mu\nu} + \mc Q^\mu u^\nu + \mc Q^\nu u^\mu + \mc T^{\mu\nu}, \\*
    J_5^\mu &= \mc N u^\mu + \mc J^\mu,
\end{align}
\end{subequations}
where
\begin{subequations} \label{Aconstitutive-kovtun}
\begin{align}
\label{AE}
\mathcal{E} &= \ve + \cfe_1 D\ve+ \ve_2 \nabla_\lambda u^\lambda+ \cfe_3 Dn_5, \\
\mathcal{P} &= P+\cfp_1 D\ve+\pi_2 \nabla_\lambda u^\lambda+\cfp_3 Dn_5, \\
\mathcal{N} &= n_5+\cfn_1 D\ve+\nu_2 \nabla_\lambda u^\lambda+\cfn_3 Dn_5, \\
\label{Q}
\mathcal{Q}^\mu &= \cfq_1 \n_\perp^\mu \ve + \theta_2 D u^\mu + \cfq_3 \n_\perp^\mu n_5 + \theta_E E^\mu + \theta_B B^\mu + \theta_\omega \omega^\mu, \\
\mathcal{J}^\mu &= \cfj_1 \n_\perp^\mu \ve + \gamma_2 D u^\mu + \cfj_3 \n_\perp^\mu n_5 + \gamma_E E^\mu + \gamma_B B^\mu + \gamma_\omega \omega^\mu, \\
\mathcal T^{\mu\nu} &= - 2 \eta \sigma^{\mu\nu}.
\end{align}
\end{subequations}
Here $\varepsilon$, $P$, and $n_5$ are the equilibrium energy density, pressure, and axial charge density, $u^\mu$ is the fluid velocity, $\omega^\mu = \frac{1}{2}\epsilon^{\mu\alpha\beta\sigma}u_\alpha \partial_\beta u_\sigma$ is the fluid vorticity, ${\sigma^{\mu\nu} = (\Delta^{\mu\alpha}\n_\alpha u^\nu + \Delta^{\nu\alpha}\n_\alpha u^\mu)/2 - \Delta^{\mu\nu}\n_\alpha u^\alpha /3}$ is the shear stress tensor, 
$D=u_\mu \nabla^\mu$, ${\nabla_\perp^\mu=\Delta^{\mu\nu}\nabla_\nu}$, and various transport parameters have been introduced. 
Note that the coefficients $\gamma_B$ and $\gamma_\omega$ encode the well-known chiral magnetic and chiral vortical effects, respectively. 

We then derive the necessary and sufficient conditions under which the equations of motion, obtained by inserting \eqref{decomp-kovtun} into the conservation laws \eqref{conservation}, are causal in the fully nonlinear regime. We obtain these conditions using standard techniques from the theory of partial differential equations tailored to the system’s characteristics \cite{Courant_and_Hilbert_book_2}. The conditions are a collection of inequalities among the transport parameters.

\begin{figure}[h]
\centering
\includegraphics[width=.7\textwidth,clip]{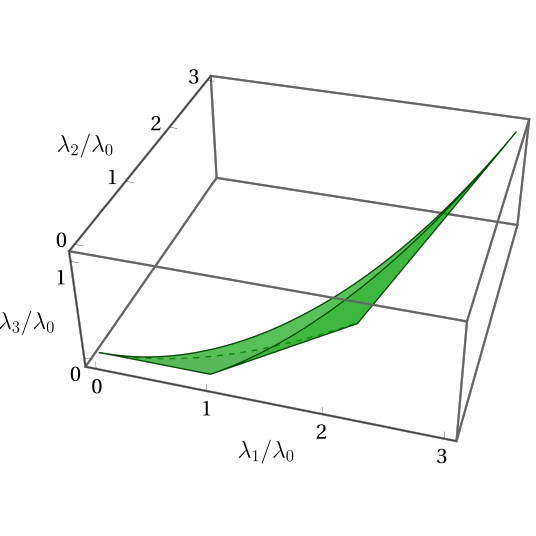}
\caption{The causal region of parameter space for first-order relativistic chiral hydrodynamics \cite{Abboud:2023hos}. The $\lambda_i$ are certain combinations of transport parameters, reported in full in \cite{Abboud:2023hos}. Provided the characteristic speed of shear waves $\eta/\theta_2$ is less than the speed of light and $\theta_\omega$ vanishes as discussed in the main text, we find that each point within the green shaded region corresponds to a causal hydrodynamic theory. In contrast, each point outside necessarily admits acausal solutions.}
\label{fig-1}       
\end{figure}

Remarkably, causality demands the heat diffusion term proportional to vorticity, i.e., $\theta_\omega \omega^\mu$ in \eqref{Q}, to vanish. To address this issue, we recall that the hydrodynamic fields $\varepsilon$, $n_5$, and $u^\mu$ do not have first-principles definitions in out of equilibrium, but rather are effective fields parameterizing the evolution of the observable fluxes $T^{\mu\nu}$ and $J^\mu$. The choice used to define the hydrodynamic fields out of equilibrium is known as a choice of hydrodynamic frame (see, e.g., \cite{Kovtun:2019hdm}). We elect to make the frame transformation, or field redefinition, given by 
\begin{equation}
  u^\mu \rightarrow u^\mu + \frac{\theta_\omega\, \omega^\mu}{\varepsilon + P}. 
\end{equation}
After inserting this transformation into \eqref{decomp-kovtun} and \eqref{Aconstitutive-kovtun} and retruncating to first order in derivatives, we obtain a theory free of the problematic term. We note that this particular definition of $u^\mu$ requires departing from the thermodynamic frame introduced in Ref.\  \cite{Jensen:2012jh}. Beyond this, only a few specific combinations of the $20$ transport parameters are required to determine causality. These combinations are denoted by $\lambda_i$, $i=0, \dots ,3$, in Fig.~\ref{fig-1}, and their explicit expressions are given in \cite{Abboud:2023hos}.

To study stability, we linearize the equations of motion around all homogeneous thermal equilibrium states in zero external fields and perform a Fourier stability analysis to identify the necessary and sufficient conditions for all perturbations to remain bounded over time. The full set of conditions is reported in Ref.\ \cite{Abboud:2023hos}. While significantly more complicated than the conditions for causality, the stability conditions are again only a set of inequalities that can be checked numerically given specific choices of the transport parameters' functional forms.

Finally, we note that our analysis reduces to ``ordinary'' non-chiral BDNK upon setting $\theta_B=\theta_\omega=\gamma_B=\gamma_\omega=0$. Therefore, we have improved upon existing results for non-chiral BDNK theory by providing the full necessary and sufficient conditions for causality and stability. Previously, only sufficient conditions were available in an explicit form.

\section{Conclusion}

Following the BDNK method, we have written the most general first-order theory of viscous chiral hydrodynamics and derived the necessary and sufficient conditions for it to be nonlinearly causal and linearly stable. These conditions are expressed in terms of inequalities that constrain the values of transport coefficients, which implies that only a specific set of hydrodynamic frames can be used. 
Within the allowed frames, it will be possible, for the first time, to consistently simulate chiral effects throughout the hydrodynamic evolution of the GQP produced in heavy-ion collisions, incorporating the dynamics of the axial and vector charge currents on the same footing as the flow of energy and momentum.

\section*{Acknowledgements}
NA and JN are partly supported by the U.S. Department of Energy, Office of Science, Office for Nuclear Physics
under Award No. DE-SC0021301 and DE-SC002386. ES has received funding from the European Union’s Horizon Europe research and innovation program under the Marie Sk\l odowska-Curie grant agreement No. 101109747.

\bibliography{refs}
\end{document}